\documentclass[a4paper,11pt]{article}
\usepackage{pos}

\title{Forward production of prompt neutrinos in the atmosphere and at high-energy colliders}
 \ShortTitle{Prompt neutrinos from the atmosphere and colliders}

\author*[a]{Yu Seon Jeong}
\author[b]{Weidong Bai}
\author[c]{Milind Diwan}
\author[d]{Maria Vittoria Garzelli}
\author[c,e,f]{Karan Kumar}
\author[g]{Mary Hall Reno}

\affiliation[a]{Chung-Ang University, High Energy Physics Center,\\
84 Dongjak-gu, Seoul, Republic of Korea}
\affiliation[b]{Sun Yat-sen University, School of Physics, \\ 
Guangzhou, Guangdong 510275, P. R. China}
\affiliation[c]{Brookhaven National Laboratory, \\
Upton, New York, USA}
\affiliation[d]{Universit\"at Hamburg, II Institut f\"ur Theoretische Physik, \\
Luruper Chaussee 149, D-22761, Hamburg, Germany}
\affiliation[e]{Stony Brook University, Department of Physics and Astronomy, \\ Stony Brook, NY 11794, USA}
\affiliation[f]{Cornell University, Department of Physics, \\
Ithaca, NY 14853, USA}
\affiliation[g]{University of Iowa, Department of Physics and Astronomy, \\
Iowa City, IA 52242, USA}

\emailAdd{yusjeong@cau.ac.kr}
\emailAdd{baiwd3@mail.sysu.edu.cn}
\emailAdd{diwan@bnl.gov}
\emailAdd{maria.vittoria.garzelli@desy.de}
\emailAdd{fk237@cornell.edu}
\emailAdd{mary-hall-reno@uiowa.edu}

\abstract{
The atmospheric neutrino flux at very high energies is dominated by prompt neutrinos, mostly contributed by the decays of charmed hadrons produced in the forward direction by cosmic ray interactions with air nuclei. Theoretical predictions of the prompt atmospheric neutrino flux have large uncertainties mainly related to charm hadron production. Prompt neutrinos can also be studied through high-energy colliders. In particular, two ongoing forward experiments and the proposed Forward Physics Facility at the LHC can detect forward prompt neutrinos.
We will present the kinematic regions relevant to the prompt atmospheric neutrino flux in terms of collider kinematic variables, the collision energy $\sqrt{s}$ and the charm hadron’s center-of-mass rapidity $y$, and discuss implications of the forward experiments at the LHC on the theoretical predictions of the prompt atmospheric neutrino flux.}

\ConferenceLogo{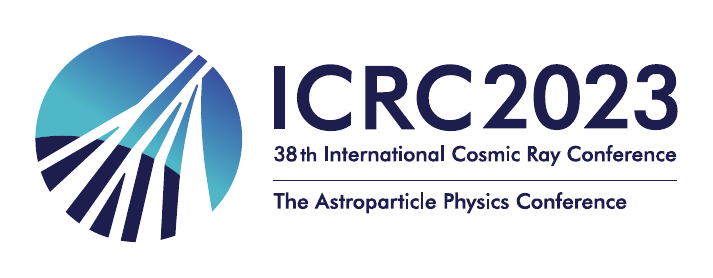}

\FullConference{%
38th International Cosmic Ray Conference (ICRC2023)\\
  26 July - 3 August, 2023\\
  Nagoya, Japan}

\begin{document}
\maketitle

\section{Introduction}

Cosmic ray interactions in the Earth's atmosphere produce a cascade of various particles, some of which decay into neutrinos, called atmospheric neutrinos.
Due to broad energy spectrum of cosmic rays, atmospheric neutrinos generated from their interactions are also distributed in a wide energy range. 
Typical particles that create atmospheric neutrinos are charged pions ($\pi^\pm$) and kaons ($K^\pm$).
Neutrinos from these light meson decays are referred to as conventional neutrinos and distributed at relatively low energies dominating the atmospheric neutrino flux up to $\sim10^5$~GeV. 
On the other hand, at very high energies,  neutrinos are also produced from heavier hadrons that contain a heavy quark, which are called prompt neutrinos and come mostly from charm hadrons.

Pions and kaons are relatively long-lived particles, and their decay lengths become longer as energy increases.
Then, they are likely to lose energy through the interactions with other particles until they decay. 
As a result, the flux of conventional atmospheric neutrinos rapidly decreases with energy. 
By comparison, the decay lengths of charm hadrons are extremely short even at high energies, therefore they immediately decay and the prompt atmospheric neutrino flux has a harder energy spectrum than the conventional flux.
Consequently, the fluxes of conventional neutrinos and prompt neutrinos cross over at a certain energy.
From various theoretical evaluations, the cross-over energy is expected to be in the range of $E_\nu \sim 10^5 - 10^6$~GeV \cite{Bhattacharya:2016jce,Zenaiev:2019ktw,Gauld:2015kvh, Garzelli:2015psa,Bhattacharya:2015jpa,Enberg:2008te,Fedynitch:2015zma,Goncalves:2017lvq,Jeong:2021vqp,Bai:2022xad} for muon neutrinos and antineutrinos. 
In this energy region, high-energy neutrinos from astrophysical sources have been observed by IceCube as a diffuse flux \cite{IceCube:2013cdw,IceCube:2020acn,IceCube:2020wum}, for which the prompt atmospheric neutrino can be the primary background. 
At present, analyses of observational data based on existing models of astrophysical neutrino fluxes indicate that the atmospheric neutrino flux can be described by only conventional neutrinos. 
Prompt atmospheric neutrinos have not been detected yet, and there are only predictions from theoretical evaluations, which have currently large uncertainties.
Two of the most important components that are responsible for large uncertainties are the incident cosmic-ray spectrum and heavy-flavor production.

Today, prompt neutrinos can be probed through high energy colliders as well.  
Over the past few years, two neutrino experiments have been prepared and installed at the LHC, FASER$\nu$ \cite{FASER:2020gpr} and SND@LHC \cite{SHiP:2020sos,SNDLHC:2022ihg}. 
The detectors are located  at a distance of 480 m from the ATLAS interaction point, one of the four proton beam collision points. They are designed to detect neutrinos produced from the $pp$ collisions and emitted into very forward direction. 
Both experiments started last year and recently reported the detection of collider neutrinos for the first time from the data collected during 2022 \cite{FASER:2023zcr,SNDLHC:2023pun}. 
These experiments will be continuously operating during the Run 3 of the LHC. 
In the meantime, a set of next-stage experiments for the High-Luminosity era of the LHC (HL-LHC) have been proposed as a collective project under the name of Forward Physics Facility (FPF) \cite{Anchordoqui:2021ghd,Feng:2022inv}. 
The FPF will include three neutrino experiments: the expanded versions of current experiments, FASER$\nu$2, AdvSND and an additional liquid Argon detector FLArE, locating the detectors at a distance of 620-685 m from the ATLAS interaction point.

At the FPF, the prompt neutrinos could be studied with very high statistics. The estimated number of neutrino interactions in FPF detectors are $\sim10^6$ for muon neutrinos and $\mathcal{O}(10^5)$ for electron neutrinos \cite{Feng:2022inv}.
The LHC at the HL-LHC stage will be run with the collision energy $\sqrt{s}= 14 {\ \rm TeV}$, which is equivalent to an energy of $\sim 10^8$~GeV in a fixed-target frame.
This energy is in a relevant region to explore astrophysical neutrinos and prompt atmospheric neutrinos. 
Therefore, measurements of prompt neutrinos and study of the heavy-flavor production through forward experiments at the LHC will help us to better understand and estimate the prompt atmospheric neutrino fluxes.
To demonstrate the relevance of the FPF for probing atmospheric neutrinos and astrophysical neutrinos, in this work we investigate the kinematic regions for prompt atmospheric neutrinos using collider variables, collision energy $\sqrt{s}$, and center-of-mass (CM) rapidity of charm hadrons $y$.

\section{Prompt atmospheric neutrino fluxes}     
Atmospheric neutrino fluxes can be evaluated using the so-called $Z$-moment method that gives an approximate solution to the coupled cascade equations for incident cosmic rays, secondary hadrons and leptons from the hadron decays.
The cascade equations describe the propagation of the high-energy particles in the atmosphere, given by 
\begin{eqnarray}
\frac{d\phi_j(E,X)}{dX}&=&-\frac{\phi_j(E,X)}{\lambda_j(E)} - \frac{\phi_j(E,X)}{\lambda^{\rm dec}_j(E,X)}
+ \sum_k S(k\to j)\,
\end{eqnarray}
with $\phi_j(E,X)$ the flux of a particle $j$ at the column depth $X$, and $\lambda_j^{(\rm dec)}$ interaction (decay) length.
The source term $S (k\to j)$ involves the particle $j$ produced by interaction or decay, and can be expressed with the energy distribution of the produced particle, $dn(k \to j)/dE$ that depends on the production process
\begin{eqnarray}
\label{eq:src}
S(k\to j) &=& \int_E^{\infty}dE ' \frac{\phi_k(E',X)}{\lambda_k(E')}
\frac{dn(k\to j;E',E)}{dE}  
\, .
\end{eqnarray}

Under the assumption $\phi_k (E',X)/\phi_k(E,X)\simeq \phi_k(E',0)/\phi_k(E,0)$, eq.~(\ref{eq:src}) can be approximated in terms of energy dependent $Z$ moment, the flux and interaction/decay length of the parent particle $k$ as below:
\begin{eqnarray}
S(k\to j) &\simeq & Z_{kj}(E)\frac{\phi_k(E,X)}{\lambda_k(E)}\, , \\
Z_{kj}(E) &\equiv& \int _E^{\infty}dE ' \frac{\phi_k^0(E')}{\phi_k^0(E)}
\frac{\lambda_k(E)}{\lambda_k(E')}
\frac{dn(k\to j;E',E)}{dE} \, .
\label{eq:zkj}
\end{eqnarray}
The resulting flux of atmospheric neutrinos can be obtained in terms of two approximate solutions of the coupled cascade equations by
\begin{equation}
\phi_\nu = \sum_h \frac{\phi_{h\to \nu}^{\rm low}\phi_{h\to \nu}^{\rm high}}{(
\phi_{h\to \nu}^{\rm low}+\phi_{h\to \nu}^{\rm high})}\ ,
\label{eq:phinu}
\end{equation}
where the two fluxes in the low-energy and high-energy limits,  $\phi_{h\to \nu}^{\rm low}$ and $\phi_{h\to \nu}^{\rm high}$ are expressed in terms of the $Z$-moments, incident cosmic ray flux $\phi_p^0$ and critical energy $\epsilon_k$ as 
\begin{eqnarray}
\label{eq:phinu_lo}
\phi_{h\to \nu}^{\rm low} &= &\sum_h \frac{Z_{ph}Z_{h\nu}}{1-Z_{pp}}\phi_p^0\, ,\\ \label{eq:phinu_hi}
\phi_{h\to \nu}^{\rm high} &= &\sum_h \frac{Z_{ph}Z_{h\nu}}{1-Z_{pp}}\frac{\ln(\Lambda_h/\Lambda_p)}
{1-\Lambda_p/\Lambda_h}\frac{\epsilon_h}{E}
\phi_p^0 \, 
\end{eqnarray}
given the effective interaction length $\Lambda_k = \lambda_k^{\rm int} / (1-Z_{kk})$. The critical energy $\epsilon_k\simeq(m_k c^2h_0/c\tau_k)$ separates the energy into low-energy and high-energy regimes.

In evaluating the atmospheric neutrino fluxes, one of the main input factors is the incident cosmic ray flux. 
A traditional parameterization is a broken power law (BPL) spectrum,
which is obtained under the assumption that the cosmic rays consist of only protons or nucleons. This is useful for comparisons with prior work and results from others \cite{Bhattacharya:2016jce,Zenaiev:2019ktw,Gauld:2015kvh, Garzelli:2015psa,Bhattacharya:2015jpa,Enberg:2008te,Fedynitch:2015zma,Goncalves:2017lvq,Jeong:2021vqp,Bai:2022xad}. 
Modern parameterizations of the cosmic ray spectrum are obtained considering different compositions and sources.
Two parameterizations most frequently used are referred to as H3p and H3a \cite{Gaisser:2011klf}, which take into account supernova remnants, galactic and extra-galactic sources for the origin of cosmic rays. The difference between the two spectra is the composition of cosmic rays from the extra-galactic origin: the H3p has only protons and the H3a has a mixed composition.

Another important factor is the heavy-flavor production cross section,  which is input into $dn(k \to j)/dE$ in eq.~\ref{eq:zkj} to evaluate the $Z$-moments for production.
This gives large uncertainties in the $Z$-moments, and eventually in the prediction of prompt neutrino flux. 
There are several approaches to evaluate the heavy-flavor production cross sections.
In this work, adopting our previous work \cite{Bai:2021ira}, we use perturbative QCD at next-to-leading order (NLO) with massive charm, QCD scales of ($\mu_R$, $\mu_F$) = (1, 2)~$m_T$ and intrinsic transverse momentum smearing $\langle k_T \rangle = 1.2~{\rm GeV}$.

 \begin{figure} 
    \centering
       \includegraphics[width=.65\textwidth]{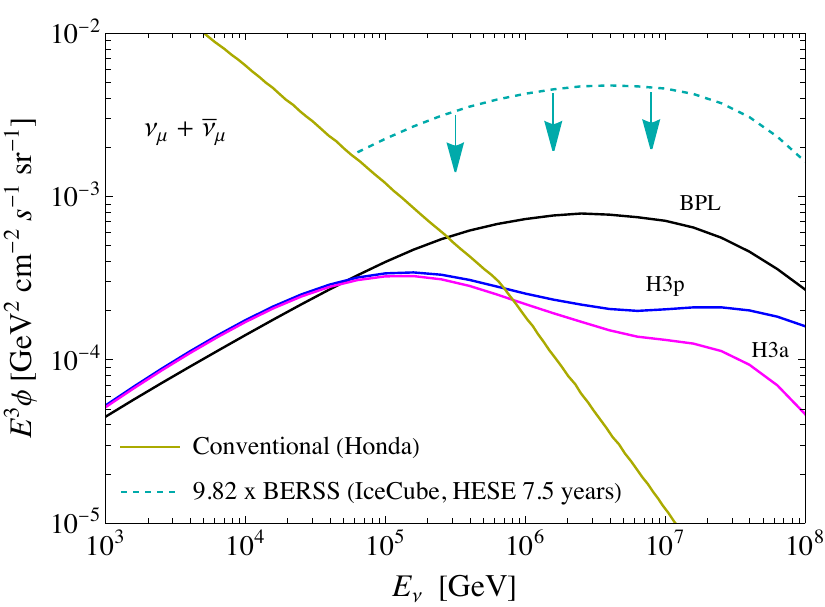}    
       \caption{The fluxes of promt atmospheric $\nu_\mu+\bar{\nu}_\mu$ from H3p, H3a and BPL cosmic-ray all-nucleon spectra. Also shown are the conventional $\nu_\mu+\bar{\nu}_\mu$ flux \cite{Honda:2006qj} and the IceCube upper limit on the prompt atmospheric neutrino flux \cite{IceCube:2020wum}. 
       The figure is taken from ref.~\cite{Bai:2022xad}.}
       \label{fig:e3phi-all}
     \end{figure}

Fig. \ref{fig:e3phi-all} shows the predictions of the prompt atmospheric $\nu_\mu+\bar{\nu}_\mu$ fluxes from charm hadron decays evaluated with different cosmic ray spectra: BPL, H3p, and H3a. 
We also present the conventional atmospheric neutrino flux \cite{Honda:2006qj} and the upper limit on  the prompt $\nu_\mu+\bar{\nu}_\mu$ flux extracted by IceCube from the analysis of 7.5 year data for high-energy starting events (HESE) \cite{IceCube:2020wum}. The upper limit is given by a scaling of the BERSS prediction \cite{Bhattacharya:2015jpa}. 
As mentioned above, one can see that the cross-over energy between the predictions of prompt and conventional atmospheric neutrino flux is between $10^5 - 10^6$~GeV.

  \begin{figure}
    \centering
       \includegraphics[width=.67\textwidth]{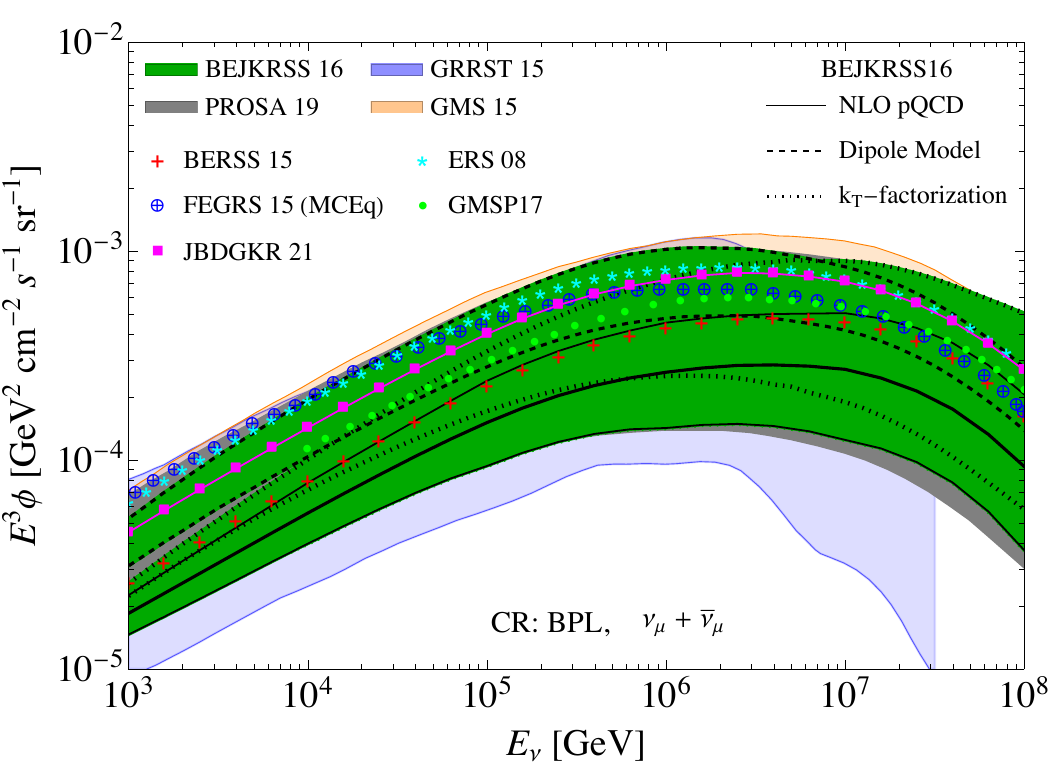}    
       \caption{Comparison of the prompt atmospheric neutrinos fluxes for $\nu_\mu + \bar{\nu}_\mu$ from refs. \cite{Bhattacharya:2016jce,Zenaiev:2019ktw,Gauld:2015kvh, Garzelli:2015psa,Bhattacharya:2015jpa,Enberg:2008te,Fedynitch:2015zma,Goncalves:2017lvq,Jeong:2021vqp,Bai:2022xad}. 
      }
       \label{fig:e3phi-unc}
     \end{figure}

In Fig.~\ref{fig:e3phi-unc}, we present some existing predictions evaluated by different groups using the BPL cosmic ray spectrum including the result from this work, referred to as JBDGKR21 \cite{Jeong:2021vqp,Bai:2022xad}. 
As shown in the figure, the uncertainty in the predictions of prompt flux is very large across the energy whereas the impact by different cosmic ray spectra appears at $E_\nu \gtrsim 10^5$~GeV. 
This uncertainty comes from various factors involved in the evaluation of the prompt flux. However, it is mostly related to the charm hadron production cross section.

\section{Connection with collider neutrinos}

\begin{figure}
    \centering
       \includegraphics[width=.49\textwidth]{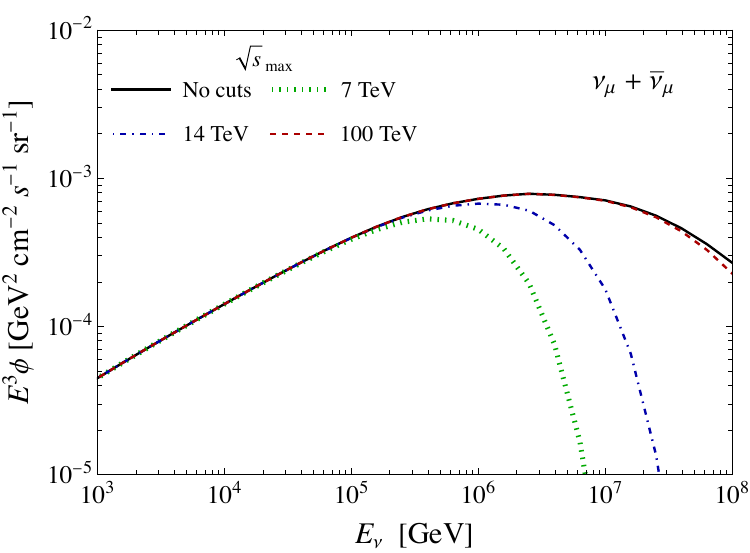}
       \includegraphics[width=.49\textwidth]{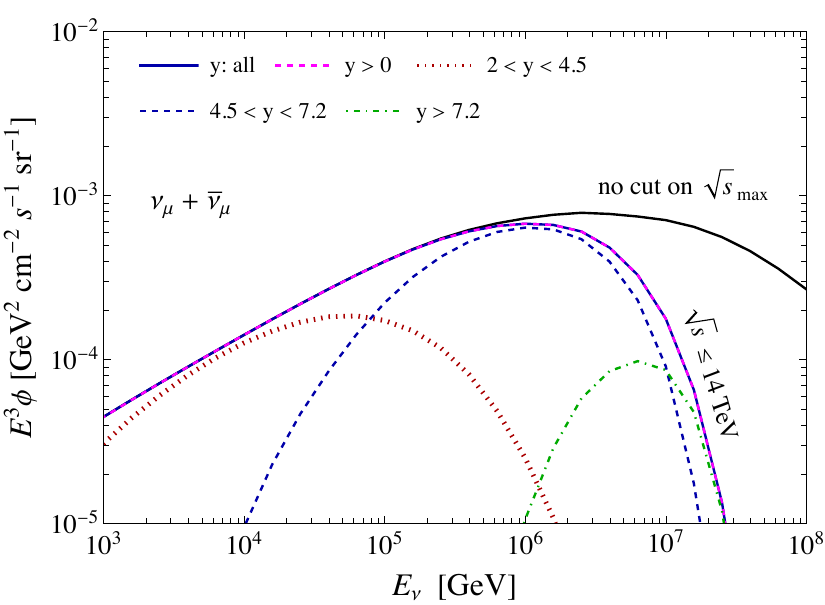}      
       \caption{The prompt flux of atmospheric $\nu_\mu+\bar{\nu}_\mu$ for different values of collision energies $\sqrt{s}$ (left) and 
       from different charm meson rapidity ranges in $pp$ collisions evaluated evaluated with $\sqrt{s}<14 {\ \rm TeV}$ (right).
       The BPL cosmic-ray spectrum is used in evaluation. The figures are taken from ref.~\cite{Bai:2022xad}.}
       \label{fig:e3phi}
     \end{figure}

We use the BPL cosmic-ray spectrum to illustrate the impact of hadronic collisions at different $\sqrt{s}$ and of charmed mesons produced in different CM rapidities on the prompt atmospheric neutrino flux.
The left panel of Fig.~\ref{fig:e3phi} shows the prompt atmospheric $\nu_\mu+\bar\nu_\mu$ fluxes for different values of the maximum CM collision energy $\sqrt{s}_{\rm max}=7$, 14 and 100 TeV with the prediction evaluated for the full range of $\sqrt{s}$.  
The first two values are the respective energies for Run 1 and HL-LHC of the LHC, and $\sqrt{s}=100$~TeV is the $pp$ collision energy considered for the Future Circular Collider (FCC). 
One can see that the maximum energy of the LHC cannot cover all the region for the prompt atmospheric neutrinos, while neutrinos from the 100 TeV collision energy contribute to most of the energy region interesting for prompt atmospheric neutrinos. Although $\sqrt{s} = 14$~TeV is equivalent to about 100 PeV in a fixed target frame, the produced neutrinos are distributed at lower energies. 
However, collisions at the LHC with this $\sqrt{s}$ still allow to cover the interesting energy region where the transition between conventional and prompt neutrinos occurs and a comparable flux of astrophysical neutrinos exists.

In the right panel of Fig.~\ref{fig:e3phi}, we show the contributions of charm hadrons produced in different CM charm hadron rapidity regions to the prompt atmospheric neutrino fluxes evaluated using the maximum collision energy $\sqrt{s}_{\rm max} = 14$~TeV. 
We divide the rapidity range into three parts: $2 < y < 4.5$, $4.5 < y < 7.2$ and $y > 7.2$.
The range of $2 < y < 4.5 $ is covered by the LHCb experiment, which is most forward region for heavy-flavor production probed at the LHC so far. 
The region of $y>7.2$ can be explored by forward experiments, both the first stage experiments (FASER$\nu$ and SND@LHC) and at the FPF. 
As shown in the figure, the prompt atmospheric neutrinos come mostly from the charm produced in the rapidity region beyond the LHCb coverage for the energies where the prompt atmospheric neutrinos are important. 
The contribution of the charm hadrons in the rapidity of $y>7.2$ is in the limited range of high energy.

\begin{figure}
    \centering
       \includegraphics[width=.48\textwidth]{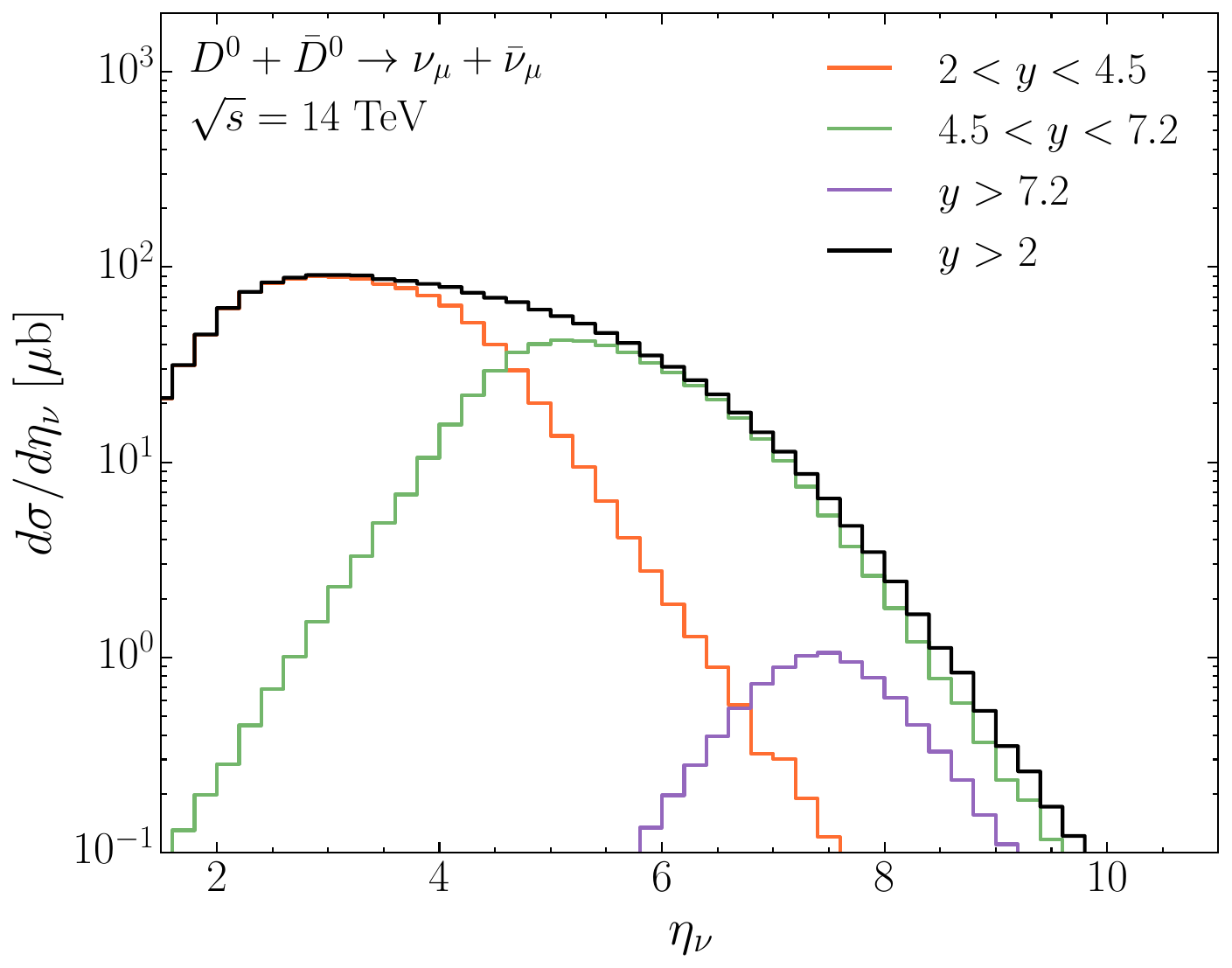}    
       \includegraphics[width=.49\textwidth]{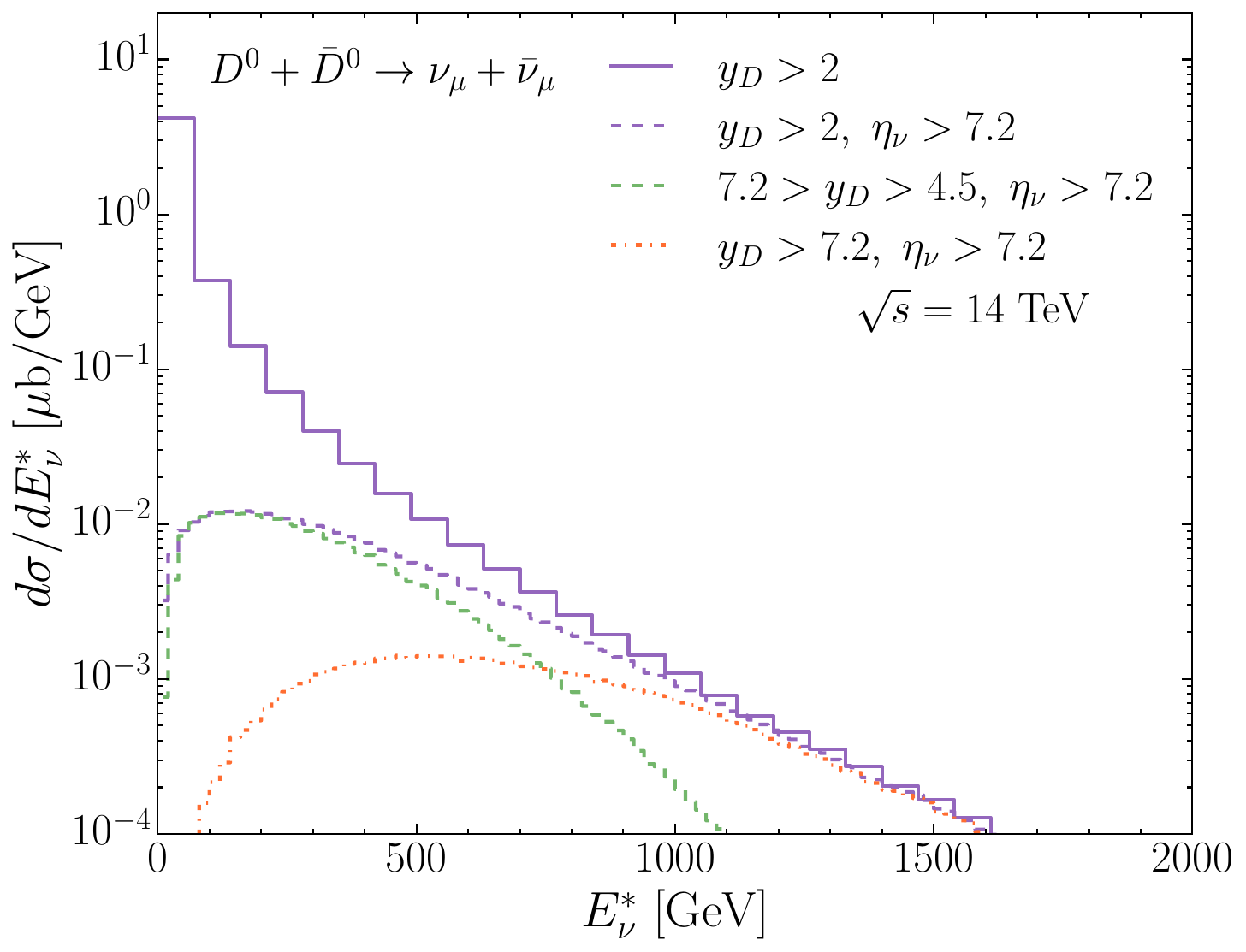}    
       \caption{Left: the neutrino rapidity ($\eta_\nu$) distribution of $\nu_\mu+\bar{\nu}_\mu$ from $D^0 + \bar{D}^0$ produced in the different charm hadron rapidity ranges from pp collisions at $\sqrt{s}=14 {\ \rm TeV}$. The figure is taken from ref.~\cite{Bai:2022xad}.
       Right: the CM frame energy distribution of $\nu_\mu+\bar{\nu}_\mu$ produced in $\eta_\nu > 7.2$ from the $D^0 + \bar{D}^0$ in different $y_D$ ranges. 
       }
       \label{fig:LHCnu}
     \end{figure}

We further focus on neutrinos that can be detected at the FPF, namely, we explore the neutrino rapidity ($\eta_\nu$) greater than 7.2.
The left panel of Fig.~\ref{fig:LHCnu} shows the $\eta_\nu$ distribution of muon neutrinos from the $D^0 + \bar{D}^0$ produced in the different charm hadron rapidity ranges from $pp$ collision at $\sqrt{s}=14$~TeV.  
This indicates that, for neutrinos that are incident into the neutrino detectors of the FPF at the LHC, charm hadrons produced in $4.5<y<7.2$ contribute more than those in $y>7.2$. 
The right panel of Fig.~\ref{fig:LHCnu} presents the CM frame energy ($E_\nu^*$) distribution of neutrinos from $D^0 + \bar{D}^0$ produced at the LHC with $\sqrt{s}=14$~TeV.
The solid histogram is for all neutrinos from $D^0 + \bar{D}^0$ in $y>2$, while the dashed histograms are for the neutrinos toward the neutrino detectors of the FPF (i.e. $\eta_\nu>7.2$) from the different $D^0+\bar{D}^0$ rapidity ranges discussed above. 
One can see that at very high CM frame energies of $E_\nu^* \gtrsim 1$~TeV, neutrinos detected at the FPF mostly come from the charm hadrons produced at the LHC in $y>7.2$. 
However, for hundreds GeV of $E^*_\nu$, most contributions to neutrinos at the FPF are predominantly from  charm hadrons with $4.5<y<7.2$, which is important region for the prompt atmospheric neutrinos. 

\section {Discussion}

We have investigated kinematic regions for prompt neutrinos produced in the atmosphere in terms of center-of-mass collision energy $\sqrt{s}$ and collider-frame rapidity of charm hadrons $y$.
Focusing on the atmospheric neutrino energy range of $10^5~{\rm GeV} < E_\nu < 10^7~{\rm GeV}$, where the prompt atmospheric neutrinos can be the main component of the atmospheric neutrino flux and play a role as an important background to the diffuse astrophysical neutrino flux, we show there is a  kinematic overlap for prompt neutrino production in the atmosphere and at the LHC. 
Although the LHC energy cannot contribute to the full energy region of prompt atmospheric neutrinos, it is high enough to cover the important energy range mentioned above. 
In the energy range of $10^5 - 10^7~{\rm GeV}$, prompt atmospheric neutrinos come mostly from the charm hadrons produced in the rapidity range of $4.5<y<7.2$, which is  beyond the coverage of the current LHC experiments that measure charm hadron production.  
However, the FPF can detect neutrinos from the decays of the charm hadrons in this rapidity region.

The prompt neutrino measurement at the FPF will help to understand charm production, constraining the parton distribution functions (PDFs) and QCD evaluations for heavy-flavor production. 
Consequently, 
it will potentially improve predictions of prompt atmospheric neutrino fluxes. 
Current analyses by IceCube with several existing models for astrophysical neutrino fluxes are compatible with zero-background of prompt neutrinos. 
We can expect that the study of prompt neutrinos at the FPF with abundant events will be able to test the assessment of the backgrounds to the astrophysical neutrino flux, which may require modification of the astrophysical neutrino flux models.
Therefore, measurements of prompt neutrinos at the FPF of the LHC will shed light on the study of astrophysical neutrinos.

\acknowledgments
This work is supported in part by U.S. Department of Energy Grants DE-SC-0010113 and DE-SC-0012704, the National Research Foundation of Korea (NRF) grant funded by the Korea government (Ministry of Science and ICT) (No. 2021R1A2C1009296) and by the German Bundesministerium f\"ur Bildung und Forschung (contract 05H21GUCCA).

%
%
%

\end{document}